\documentclass[twocolumn,showpacs,preprintnumbers,amsmath,amssymb]{revtex4}

\usepackage{graphicx}
\usepackage{dcolumn}
\usepackage{bm}

\begin{document}

\preprint{APS/123-QED}

\title{Randomness-induced quantum spin liquid on honeycomb lattice}

\author{H. Yamaguchi$^{1}$, M. Okada$^1$, Y. Kono$^2$, S. Kittaka$^2$, T. Sakakibara$^2$, T. Okabe$^1$, Y. Iwasaki$^1$, and Y. Hosokoshi$^1$}
\affiliation{
$^1$Department of Physical Science, Osaka Prefecture University, Osaka 599-8531, Japan \\ 
$^2$Institute for Solid State Physics, the University of Tokyo, Chiba 277-8581, Japan \\
}

Second institution and/or address\\
This line break forced

\date{\today}

\begin{abstract}
We present a quantu spin liquid state in a spin-1/2 honeycomb lattice with randomness in the exchange interaction. 
That is, we successfully introduce randomness into the organic radial-based complex and realize a random-singlet (RS) state. 
All magnetic and thermodynamic experimental results indicate the liquid-like behaviors, which are consistent with those expected in the RS state.
These results demonstrate that the randomness or inhomogeneity in the actual systems stabilize the RS state and yield liquid-like behavior. 
\end{abstract}

\maketitle
A quantum spin liquid (QSL) is one of the fascinating ground states encountered in the field of condensed matter physics.
In a QSL, enhanced quantum fluctuations in strongly correlated spins prevent magnetic ordering, inducing the formation of a disordered state that exhibits liquid-like spin behavior. 
Anderson proposed the resonating valence bond as a possible QSL state in the $S$=1/2 frustrated triangular lattice~\cite{RVB}.
Following considerable experimental research, several candidate materials have since been reported. 
For example, the organic salts $\kappa-$(BEDT-TTF)$_2$Cu$_2$(CN)$_3$~\cite{bedt1,bedt2,bedt3} and EtMe$_3$Sb[Pd(dmit)$_2$]$_2$~\cite{dmit1,dmit2,dmit3}, which form $S$=1/2 Heisenberg antiferromagnetic (AF) triangular lattices are known to be promising candidates.
These materials have no magnetic order down to very low temperatures and exhibit gapless (or nearly gapless) behaviors.  
However, theoretical research has established that the ground state of the triangular lattice is an AF ordered state~\cite{120order1,120order2}.   
Thus, the true origin of the liquid-like behavior observed in these organic salts remains an open question.

Subsequent theoretical studies on the triangular lattice with the liquid-like behavior have revealed several mechanisms that may stabilize a QSL state. 
The numerical calculations employed in those investigations incorporate additional effects that are neglected in the simplest model with nearest-neighbor bilinear coupling~\cite{QSL_cal1,QSL_cal2,QSL_cal3,QSL_cal4,QSL_cal5}. 
Meanwhile, it has been noted that the inhomogeneity in the actual systems, which causes spatially random exchange coupling (bond-randomness), may be essential for the observed liquid-like behavior~\cite{RS_sankaku,RS_sankaku_kagome}. 
As regards organic salts, strong coupling between the spin and electric polarization at each spin site and its importance to the liquid-like behavior have been noted~\cite{pola1,pola2}.
In the low-temperature region, random freezing of the electiric polarization is indicated by the glassy response in the dielectri properties~\cite{charge1, charge2}.  
Accordingly, the spin-density distribution at each lattice site should be also randomly freezing and causes the bond-randomness.

Recent numerical analysis of the bond-randomness effect on an $S$=1/2 Heisenberg AF triangular lattice has revealed that sufficiently strong randomness stabilizes a gapless QSL state~\cite{RS_sankaku,RS_sankaku_kagome}.
Such a randomness-induced QSL, in which spin-singlet dimers of varying strengths are formed in a spatially random manner, corresponds to a so-called random-singlet (RS) or valence-bond glass (VBG)~\cite{RS1,RS2,VBG1,VBG2}. 
Because the exchange interactions are randomly distributed, the binding energy of the singlet dimers has a wide distribution, yielding gapless behavior.  
The RS (or VBG) state is expected to exhibit liquid-like behaviors characterized by a $T$-linear specific heat, a gapless susceptibility with an intrinsic Curie tail, a near-linear magnetization curve~\cite{RS_sankaku}.
Indeed, many of the experimentally observed liquid-like behaviors of the triangular organic salts are explained by the RS picture.
Further numerical studies suggested that bond-randomness on kagome and honeycomb lattices also stabilizes the RS state~\cite{RS_kagome,RS_honeycomb}.
In the case of an $S$=1/2 Heisenberg AF honeycomb lattice, which is our focus in this letter, the ground-state phase diagram for the bond-randomness versus frustration is investigated~\cite{RS_honeycomb}. 
It is suggested that liquid-like behavior can be realized even in the case of very weak frustration.  
Although some honeycomb-lattice-based compounds have recently been reported to exhibit liquid-like behavior, their exact lattice systems have not been clarified~\cite{BaNi_honeycomb, BaCu_honeycomb}. 

In this letter, we present a new $S$=1/2 Heisenberg AF honeycomb lattice composed of three dominant interactions, as shown in Fig. 1(a).
Those interactions are designed to have bond-randomness with a weak additional AF interaction $J_{\rm{4}}$ inducing frustration in the lattice.
Our experimental results regarding the magnetic and thermodynamic properties indicate the realization of the RS state, as schematically shown in Fig. 1(b).

\begin{figure*}[t]
\begin{center}
\includegraphics[width=36pc]{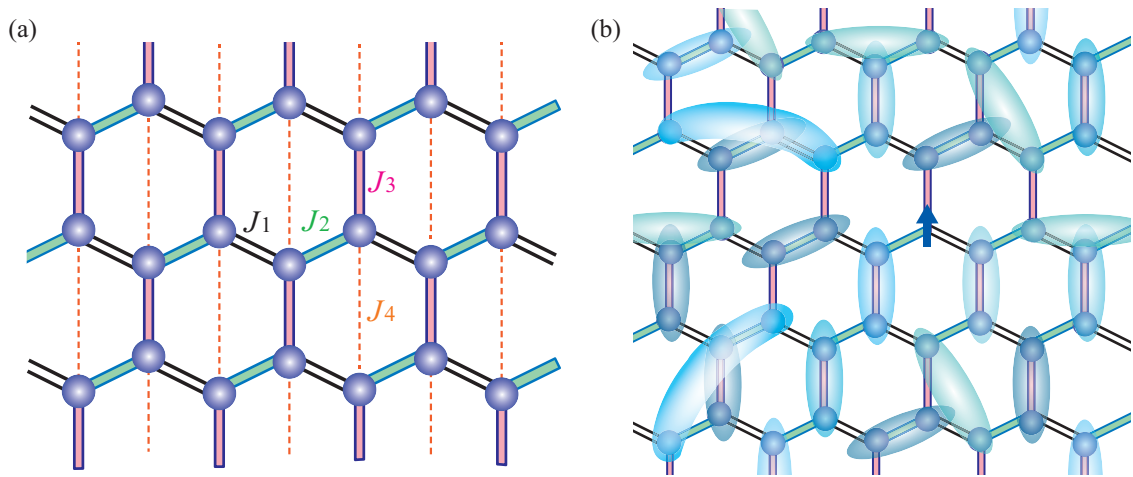}
\caption{(color online) Honeycomb lattice of Zn(hfac)$_2$(A$_x$B$_{1-x}$) (a) $S$=1/2 honeycomb lattice composed of one ferromagnetic interaction  $J_{\rm{1}}$ and two AF interactions $J_{\rm{2}}$ and $J_{\rm{3}}$ in the $ac$-plane. (b) RS state in examined honeycomb lattice. Entangled spin-singlet dimers are indicated by ovals that cover two lattice sites. The dimers can be formed in a spatially random manner, not only between neighboring sites, but also between distant sites through higher-order interactions. The arrow indicates an unpaired“orphan”spin.}\label{f1}
\end{center}
\end{figure*}

The crystal structure was determined on the basis of intensity data collected using a Rigaku AFC-8R Mercury CCD RA-Micro7 diffractometer with a Japan Thermal Engineering XR-HR10K. 
The magnetizations were measured down to approximately 80 mK using a commercial SQUID magnetometer (MPMS-XL, Quantum Design) and a capacitive Faraday magnetometer.
The experimental results were corrected for diamagnetic contributions (-4.2${\times}$ $10^{-4}$ emu mol$^{-1}$ for $x$=0.64 and -4.0${\times}$ $10^{-4}$ emu mol$^{-1}$ for $x$=0.79), which were determined to become almost $\chi T$ = const. above approximately 200 K, and close to the value calculated using Pascal's method.
The specific heat was measured using a hand-made apparatus and a standard adiabatic heat-pulse method down to $\sim$0.1 K . 
Considering the isotropic nature of organic radical systems, all experiments were performed using small randomly oriented single crystals.


We have developed verdazyl radical systems with flexible molecular orbitals (MOs) that enable tuning of the intermolecular magnetic interactions through molecular design~\cite{3Cl4FV,3ladders,26Cl2V}. 
Recently, we succeeded in combining the verdazyl radicals with 3$d$ transition metals to obtain a molecular-based complex~\cite{Znhfac, Mnhfac}.
In this study, we utilized a new verdazyl-based complex Zn(hfac)$_2$(A$_x$B$_{1-x}$), where hfac represents 1,1,1,5,5,5-hexafluoroacetylacetonate, and A and B equivalent to regioisomers of verdazyl radical.   
It is essential for introduction of randomness that the rotational degrees of freedom of verdazyl radical disappear owing to the coordination to Zn(hfac)$_2$, as shown in Fig. 2(a). 
Accordingly, two different regioisomers, labeled A-type ($x$) and B-type (1-$x$), arise and randomly align in the crystal, yielding randomness of the intermolecular exchange interactions.

\begin{figure*}[t]
\begin{center}
\includegraphics[width=36pc]{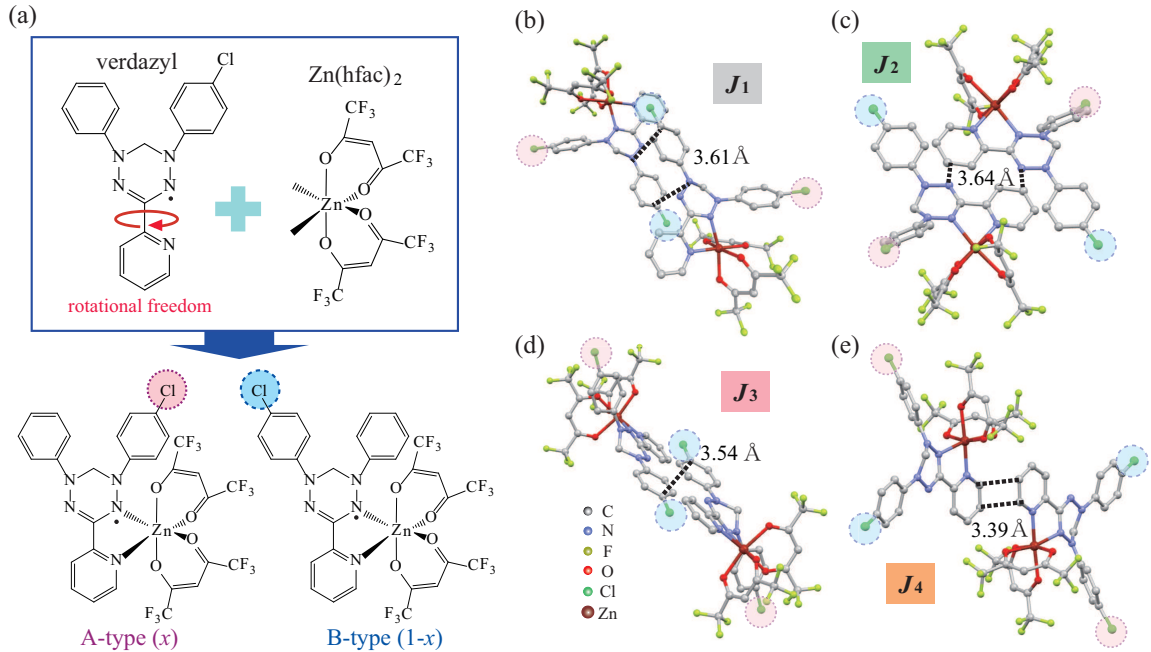}
\caption{(color online) (a) Molecular structures of verdazyl radical, Zn(hfac)$_2$, and combined complex Zn(hfac)$_2$(A$_x$B$_{1-x}$) with two regioisomers, i.e., A- ($x$) and B-type (1-$x$), where $x$ represents Cl occupancy for the A-type in the crystal. Molecular pairs associated with the magnetic interactions (b) $J_{1}$, (c) $J_{2}$, (d) $J_{3}$, and (e) $J_{4}$. The broken lines indicate C-N and C-C short contacts. Each molecular pair is related by inversion symmetry and has three pair formation patterns, A-A, A-B (=B-A), and B-B. }\label{f2}
\end{center}
\end{figure*}

The randomness effect reaches maximum at $x$=0.5, where the numbers of A- and B-type molecules are identical. 
Some numerical inequalities exist depending on the conditions of the solution used in the complex-forming reaction (see Supplementary), and the actual crystals have slightly large $x$ values.
Here, we successfully synthesized two different single crystals with x=0.64 and 0.79.
The crystallographic parameters were determined at room temperature and 25 K for both crystals (Supplementary Table S1).
Only slight differences were observed between the results for $x$=0.64 and 0.79.
Note that, because this investigation focused on the low-temperature magnetic properties, the crystallographic data at 25 K are discussed hereafter.
The crystallographic parameters at 25 K for $x$=0.64 were as follows: monoclinic, space group $P2_{1}/n$, $a$ =  9.010(3) $\rm{\AA}$, $b$ = 31.640(11) $\rm{\AA}$, $c$ = 10.902(4) $\rm{\AA}$, $V$ = 3107.8(19)  $\rm{\AA}^3$, $Z$ = 4. 
We performed the MO calculations and found three types of dominant interactions, i.e., $J_{\rm{1}}$, $J_{\rm{2}}$, and $J_{\rm{3}}$, and an additional weak interaction $J_{\rm{4}}$, as shown in Fig. 1(a).
The molecular pairs associated with those exchange interactions are all related by inversion symmetry, as shown in Fig. 2(b)-(e).


Considering the A- and B-type molecule combination with the inversion center between them, each interaction has three pair formation patterns, i.e., A-A, A-B (=B-A), and B-B.
We focused on the $x$ = 0.64 crystal as it has a stronger randomness effect
The A-A, A-B, and B-B pairs have possibilities of $x^2$ = 0.41, $2x(1-x)$= 0.46, $(1-x)^2$ = 0.13, respectively.
The exchange interactions $J_{i}/k_{\rm{B}}$ = $\{$A-A, A-B, B-B$\}$ ($i$ = 1-4; $k_B$ is the Boltzmann constant) were evaluated as $J_{1}/k_{\rm{B}}$ = $\{$$-$9.5 K , $-$14.5 K, $-$15.8 K$\}$, $J_{2}/k_{\rm{B}}$ = $\{$7.2 K , 7.1 K, 6.9 K$\}$, and $J_{3}/k_{\rm{B}}$ = $\{$3.9 K , 7.8 K, 10.0 K$\}$; these terms are defined in the Heisenberg spin Hamiltonian given by $\mathcal {H} = J_{n}{\sum^{}_{<i,j>}}\textbf{{\textit S}}_{i}{\cdot}\textbf{{\textit S}}_{j}$, where $\sum_{<i,j>}$ denotes the sum over the corresponding spin pairs. 
As the molecular-orbital overlaps for $J_{1}$ and $J_{3}$ are strongly related to the phenyl rings with randomly distributed Cl atoms, their values are highly dependent on the pair formation.
 

\begin{figure*}[t]
\begin{center}
\includegraphics[width=40pc]{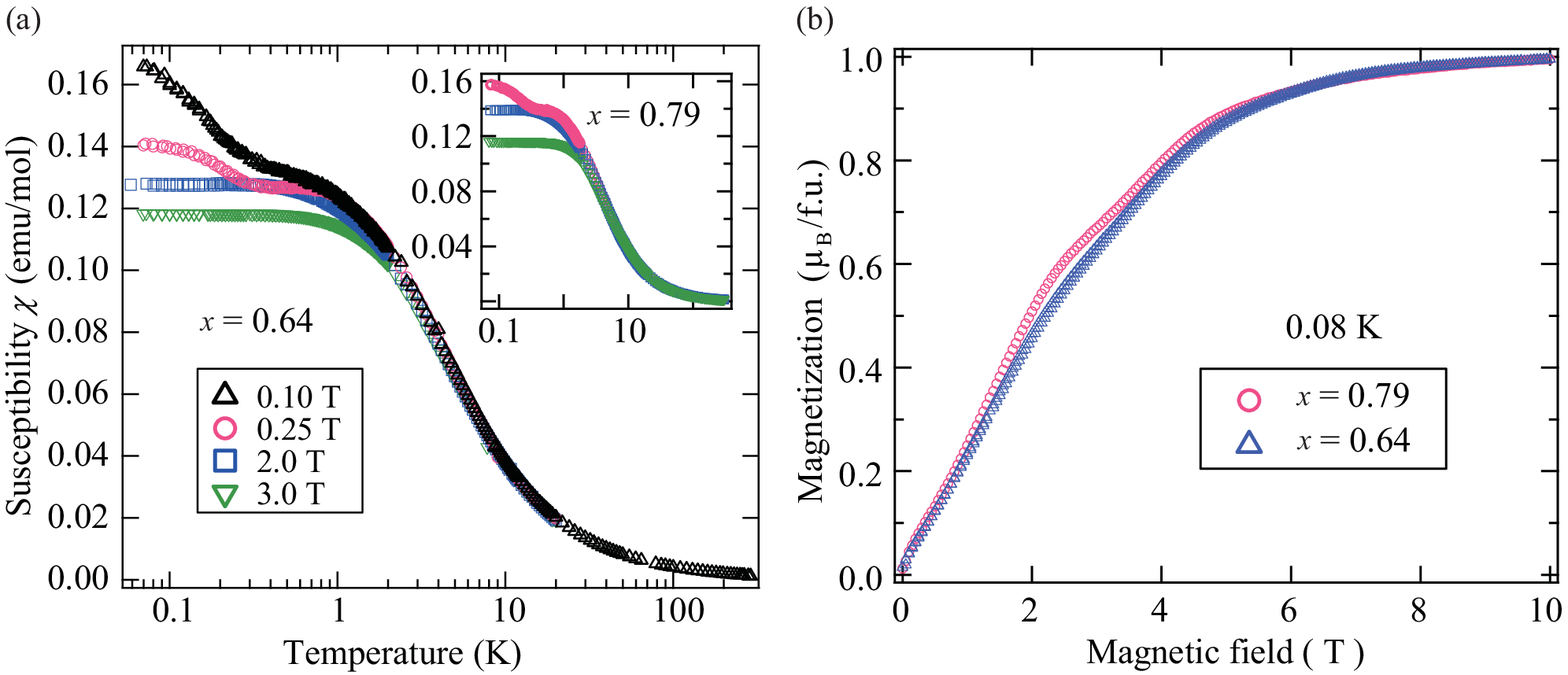}
\caption{(color online) (a) Temperature dependence of magnetic susceptibility ($\chi$ = $M/H$) for $x$=0.64 and 0.79 (inset). (b) Magnetization curve at 0.08 K for $x$=0.64 and 0.79 K. }\label{f3}
\end{center}
\end{figure*}

\begin{figure}[t]
\begin{center}
\includegraphics[width=18pc]{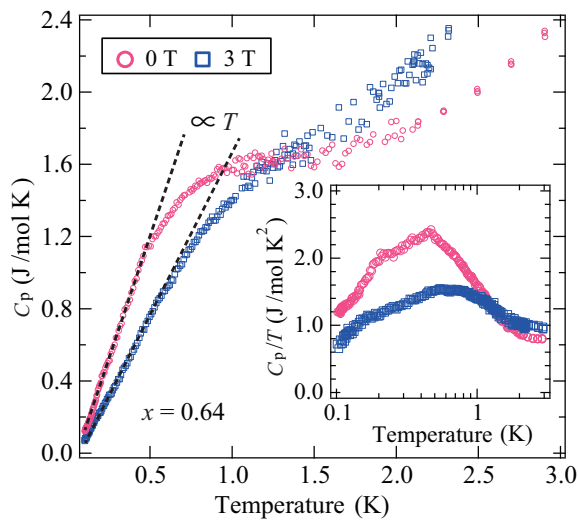}
\caption{(color online) Low-temperature specific heat of Zn(hfac)$_2$(A$_x$B$_{1-x}$) at 0 and 3 T for $x$ = 0.64. 
The broken lines indicate the $T$-linear behavior. Inset: Corresponding $C_{\rm{p}}$/$T$.}\label{f4}
\end{center}
\end{figure}

Figure 3(a) shows the temperature dependence of the magnetic susceptibilities ($\chi$ = $M/H$) for various magnetic fields.
Above 50 K, the behavior follows the Curie-Weiss law, and the Weiss temperatures are estimated to be ${\theta}_{w}$ $\approx$+2.8 K and +1.9 K for $x$=0.64 and 0.79, respectively. The ${\theta}_{w}$ sign indicates the dominant contribution of the ferromagnetic $J_{1}$, and the small absolute values indicate weak internal fields due to competition between the ferromagnetic and AF interactions.
Although we performed field-cooled and zero-field-cooled measurements to examine spin-freezing for $x$=0.64, no distinguishable differences were found, as shown in Fig. 3(a).
Below 0.25 T, a shoulder is apparent at approximately 1 K, along with gapless behavior with a Curie tail in the lower temperature region, for both $x$ = 0.64 and 0.79.
This shoulder indicates development of AF correlations forming spin-singlet dimers, and the Curie-like diverging components indicate a small fraction of free spins owing to some unpaired“orphan” spins, as illustrated in Fig. 1(b). 
The appearance of the small free-spin fraction generating Curie-like low-temperature $\chi$ is indeed expected in the RS picture~\cite{RS_sankaku,RS_kagome,RS_honeycomb}.

The magnetization curves at the lowest temperature of 0.08 K also indicate gapless behaviors, as shown in Fig. 3(b).
Small paramagnetic contributions given by the Brillouin function appear in the low-field region below $\sim$0.5 T for both $x$=0.64 and 0.79 K, which correspond to the Curie tail in $\chi$ and are evaluated to be approximately 3 
The entire magnetization curve for $x$ = 0.64 exhibits the near-linear behavior expected for the RS state~\cite{RS_sankaku}, whereas that for $x$ = 0.79 exhibits bending at approximately 3.5 T.
Such bending was also observed in the magnetization curve calculated for the RS state near the saturation field~\cite{RS_sankaku}.
In general, there is a sharp change in the magnetization curve at the saturation field for non-randomness phase.
By introducing bond-randomness, the magnetization curve near the saturation field becomes gradual, originating from the widely distributed binding energy of the singlet dimers.
That is, bending appears when the randomness is small.
When the randomness is increased, the bending eventually becomes almost linear.
As the randomness increases when $x$ approaches 0.5, our results are consistent with the theoretical prediction. 

The temperature dependence of the specific heat, $C_{\rm{p}}$, for $x$ = 0.64 is shown in Fig. 4. 
The magnetic contributions are expected to be dominant in the low-temperature regions considered here, and nuclear Schottky contributions are subtracted assuming estimation from the nuclear spins, 2.403${\times}$ $10^{-4}$$H^2$/$T^2$.
In the low-temperature regions below 0.5 K, a clear gapless $T$-linear behavior was observed, $C_{\rm{p}}{\simeq}{\gamma}T$, which is expected for the RS state. 
This $T$-linear behavior is robust against an applied magnetic field and appears even under a high-magnetic field near the saturation field.
We also found a broad hump structure in the temperature dependence of $C_{\rm{p}}$/$T$ (Fig. 4, inset).
Note that similar broad hump structures have also been observed for the $C_{\rm{p}}$/$T$ of organic triangular salts, in which the broad hump is considered to be a crossover to QSL state~\cite{bedt2,dmit3}.  
We roughly evaluated ${\gamma}$ from the $C_{\rm{p}}$/$T$ values at the lowest temperatures and obtained 1.20 and 0.72 J/mol K$^2$ at 0 and 3 T, respectively. 
From numerical analysis of the RS state, we thus deduced that the $T$-linear term of $C_{\rm{p}}$ is strongly dependent on the fundamental ground state without randomness and, also, on the degree of introduced randomness~\cite{RS_sankaku,RS_kagome,RS_honeycomb}.
The obtained ${\gamma}$-values are somewhat large, but do not differ significantly from the calculations.

In the honeycomb lattice, a relatively small lattice distortion can induce a disordered gapped phase (even in the non-frustrated case) owing to strong quantum fluctuation~\cite{honeycomb_gap1,honeycomb_gap2}. 
From the theoretical analysis, it is deduced that the introduction of bond-randomness into the gapped phases is more effective for RS state formation than introduction into the gapless ordered phase~\cite{RS_sankaku,RS_kagome,RS_honeycomb}.
Therefore, in the present model, the lattice distortion as well as the weak frustrated interaction should enhance the bond-randomness effect, inducing formation of the RS state.
Indeed, almost all magnetic and thermodynamic properties of the present model are consistent with the expected beheviors in the RS state.

In summary, we have succeeded in synthesizing single crystals of the verdazyl-based complex Zn(hfac)$_2$(A$_x$B$_{1-x}$). 
Two different regioisomers, A-type ($x$) and B-type (1-$x$), arise and randomly align in the crystal, yielding randomness of the intermolecular exchange interactions. 
$Ab$ $initio$ MO calculations indicate the formation of the $S$=1/2 Heisenberg AF honeycomb lattice composed of three dominant interactions, and there is a weak additional AF interaction inducing frustration in the lattice.
All magnetic and thermodynamic experimental results indicate the liquid-like behaviors, which are consistent with those expected in the RS state,  
These results demonstrate that the randomness or inhomogeneity in the actual systems stabilize the RS state and yield liquid-like behavior.
Furthermore, our method to introduce a bond-randomness into spin lattices enable further investigations on the randomness-induced QSL in other lattice systems.

We thank T. Kawakami for valuable discussions. This research was partly supported by KAKENHI (No. 17H04850, No. 15H03682, and No. 15H03695). Part of this work was performed as a joint-research program involving the Institute for Solid State Physics (ISSP), the University of Tokyo, and the Institute for Molecular Science.


\end{document}